\documentstyle[editedvolume,psfig]{crckapb}
\def\kmse{{km\thinspace s$^{-1}$}\ }              
\def\kms{{km\thinspace s$^{-1}$}}
\def\13co{$^{13}$CO}
\def\h2{H$_2$}

\def\b{$b$}

\def\deg{\ifmmode^\circ\else$^\circ$\fi}
\def\dege{{\ifmmode^\circ\else$^\circ$\fi}\ }
\def\solar{\ifmmode _{\mathord\odot}\else$_{\mathord\odot}$\fi}
\def\sun{$_\odot$}

\def\msun{M\sun}

\def\msune{{M\sun}\ }

\def\hper{\ifmmode \rlap.^{h} \else $\rlap{.}^h$\fi}
\def\mper{\ifmmode \rlap.^{m} \else $\rlap{.}^m$\fi}
\def\sper{\ifmmode \rlap.^{s} \else $\rlap{.}^s$ \fi}
\def\degper{\ifmmode \rlap.^{\circ} \else $\rlap{.}^{\circ} $\fi}
\def\arcmper{\ifmmode \rlap.{' } \else $\rlap{.}' $\fi}
\def\arcsper{\ifmmode \rlap.{'' } \else $\rlap{.}'' $\fi}
\def\cc{cm$^{-3}$}
\def\cce{cm$^{-3}$\ }
\def\c2{cm$^{-2}$}

\def\lapprox{$_<\atop{^\sim}$}  
\def\apjs{{ApJS},\ }      
\def\apj{{ApJ},\ }       
\def\mn{{MNRAS},\ }               
\def\aa{{A\&A},\ }       
\def\aj{AJ, \rm}
\def\m12{\magnification=1200}

\def\deg{\ifmmode^\circ\else$^\circ$\fi}              
\def\arcs{\ifmmode {'' }\else $'' $\fi}           
\def\arcm{\ifmmode {' }\else $' $\fi}             

\def\lapprox{$_<\atop{^\sim}$}  

\begin{opening}
\title{High Velocity Clouds: The Missing Link?}

\author{Leo Blitz}
\institute{University of California, Berkeley}
\author{David N. Spergel}
\institute{Princeton University}
\author{Peter J. Teuben}
\institute{University of Maryland}
\author{Dap Hartmann}
\institute{University of Cologne}

\end{opening}

\runningtitle{HVCs: The Missing Link}
\begin{document}

\begin{abstract}
Hierarchical structure formation models predict the existence
of large numbers of low velocity dispersion dark halos.  Galaxy
surveys
find far fewer galaxies than predicted by analytical estimates
and numerical simulations. In this paper,
we suggest that these dark halos are not missing,
but  have been merely misplaced in the galactic astronomy section
of the journals: they are the High Velocity
Clouds (HVCs).   We review the predictions of our model for the Local
Group origin of the HVCs and
its implications for the formation and the evolution 
of our Galaxy.  We describe  recent observations that
confirm many of earlier predictions and discuss future
tests of the model.  
\end{abstract}

\section{The Missing Galaxy Problem}

A generic prediction of hierarchical structure formation models
is the existence of large numbers of low mass halos.  The
Press-Schechter formalism (Press \& Schechter 1974)
predicts that the galaxy
mass function,
\begin{equation}
n(M) \propto M^\alpha \exp(-M/M_*)
\end{equation}
has a steep faint end slope, $\alpha \simeq -2$.  
Numerical simulations (Efstathiou et al. 1988; 
Gelb \& Bertschinger 1994, Klypin et al. 1999)
are consistent with the Press-Schechter approach: they also
predict copious low mass halos.  

Most galaxy surveys, however, do not seem to find large numbers
of low luminosity, low velocity dispersion galaxies.  Loveday (1998)
summarizes a number of recent field surveys that
find a faint end slope in the range $-1.2 \simeq \alpha \simeq -0.7$.
Groups have similar ``flat" galaxy luminosity functions with slopes
typically $\sim -1$  (Muriel, Valotto \& Lambas 1998).
While surveys that reach lower
surface brightness limits find more dwarf galaxies (Bothun,
Impey \& McGaugh 1997, Dalcanton et al. 1998), even the inclusion
of these systems does not appear to increase the faint end
slope enough to reconcile theory and observation.

In our own Local Group, where the galaxy
inventory is thought to be essentially complete, the discrepancy  is even more
severe. 
Simulations at the appropriate
scale suggest that the Local Group should contain roughly
1000 objects with velocity dispersions larger than 10 \kms (Klypin et al. 1999).
Observers however have only been able to
find $\sim 30$ galaxies in the Local Group (Mateo 1998).

Where are the missing
dark halos? There is either something wrong
with hierarchical structure formation, the numerical simulations,
or there are a host of unidentified bound systems in the Local Group.

What are the likely properties of
these low velocity dispersion halos?
Star formation is  likely to be inefficient in these
low luminosity systems because the cosmological ultraviolet
background can prevent or at least delay the formation
of atomic and molecular hydrogen (Babul \& Rees 1992;
Kepner, Babul \& Spergel 1997; Barkana \& Loeb 1999).  If these
missing galaxies
have not formed stars, they likely persist as small bound
objects containing mostly ionized hydrogen and possibly a handful
of stars.  These dark halos may be the High Velocity Clouds (HVCs).

The HVCs are clouds of atomic hydrogen detected
primarily by means of their 21-cm emission that cannot be in circular
rotation about the Galactic Center.  Because they are largely found at
high Galactic latitude, and because the HI layer of the Milky Way is
so thin, the characteristic distance to the HVCs can, in principle, be
anywhere between several hundred parsecs and 1 Mpc. In this paper, we
will argue that the evidence points to a Local Group
origin for the HVCs.  Oort's (1964) original idea that the clouds
represent infall
onto the Milky Way was abandoned long ago because of the discovery
of HVCs with positive galactocentric velocities.  Nevertheless, we
will
argue that his insight that the HVCs represent the unaccreted remnants
of galaxy formation is largely correct.  We will also offer some
speculations on the implications of the Local Group origin.  A fuller,
more detailed account of the arguments presented in this paper may
be found in Blitz et al. (1999a).  Some of the
arguments in  this article have
already appeared as Blitz et al. (1999b).

\section{Evidence for a Local Group Origin}

\subsection{Kinematic Data}

The flux from HVCs is less than 10$^{-4}$ of the normal Galactic
emission.
In the outermost parts of the Milky Way, a longitude-velocity plot of
the normal Galactic emission shows that the contours are very nearly
sinusoidal.  The radial velocity, $v_r$ of the gas in circular orbit
around the
center obeys:

\begin{equation}
v_r = \Theta(R_{\rm O}/R - 1)\ sinl\ cosb     
\end{equation}

\noindent
where $\Theta$ is the circular velocity of a gas parcel at a
distance $R$ from the Galactic Center, and where $R_{\rm O}$ is the
distance of
the Sun from the Center.  Thus, for gas close to the plane (cos$b$
$\simeq$ 1), sinusoidal profiles for a flat rotation curve are
indicative of gas at constant radius.  Conversely, the galactocentric
distance of gas along a sinusoid can be inferred from its maximum
velocity (sin$l$\thinspace cos$b$ = 1).  Longitude-velocity plots from the
Leiden-Dwingeloo survey (Hartmann \& Burton 1997) exhibit HI emission
with maximum velocities of about $\pm$ 170 \kmse (Blitz et al.  1999),
implying that for a flat rotation curve, the HI disk of the Milky Way
extends to a distance of $\simeq$ 37 kpc from the Galactic Center.

Although most of the HVCs are far from the Galactic plane, one large
cloud known as Complex  H (after Aad Hulsbosch who has spent years
observing and categorizing the HVCs), lies directly in the plane and
is
shown in Figure 1.  The velocity centroid of this cloud is --194
\kms.  If it were within the HI disk, the velocity difference between
this cloud and the gas in normal galactic rotation would be between
30--200 \kmse giving rise to a very large region of highly shocked gas
with an energy of $\simeq$ 10$^{54}$  ergs, nearly independent of
distance over a large portion of the disk.  This much energy would
give
rise to a region of disturbed gas in the HI disk 
at least 25 degrees in extent 
as well as giving rise to strong H$\alpha$ and perhaps
x-ray emission, but none of these is evident.  Cloud H must therefore
lie beyond the HI disk at a distance of at least 40 kpc from the
center.  If it is at a distance of 50 kpc, the cloud has a diameter of
20 kpc, and an HI mass of 9 $\times 10^7$ \msun, a huge HI cloud
comparable in mass to a dwarf spheroidal galaxy.  If this cloud is
typical of the others, then the typical distance of the HVCs will be
$\simeq$ 25 times larger (since the median angular size of the HVCs is
$\simeq$ 25 times smaller) and the reservoir of HI locked up in the
HVCs is enough to make a Milky Way-sized galaxy.

\begin{figure} [!htb]
\centerline{\psfig{figure=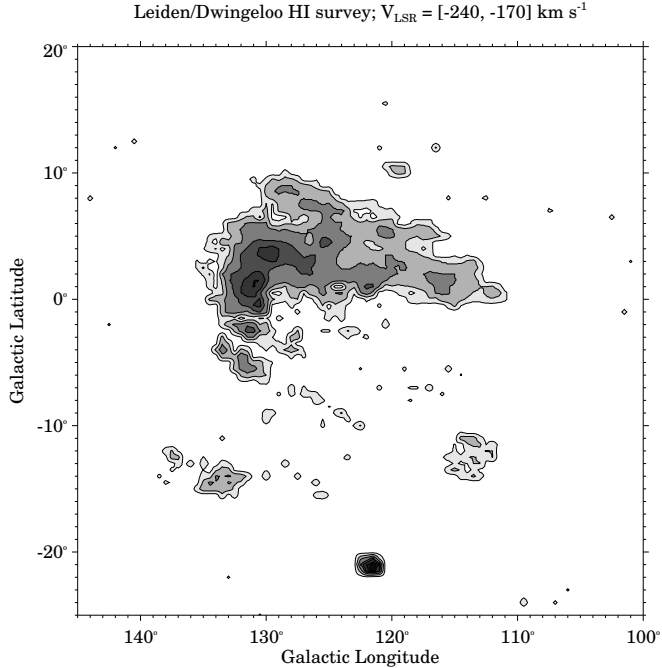,height=5in,angle=0,silent=1}}
\vskip -0.6in
\caption{HI emission from Complex H integrated over the velocity range
$-240 \leq~ {\rm V}_{\rm LSR} \leq -170$ \kms, effectively excluding the
conventional Galactic gaseous disk. 
The bright object at $l = 122\deg$, $b = -21\deg$ represents
the portion of M31 emitting within the chosen velocity range. 
[from Blitz et al. (1999a)]} 
\end{figure}

The velocity dispersion and the velocity centroid of the cloud
ensemble
can also be used to determine what the most appropriate inertial frame
of
reference is.  A non-inertial frame always gives rise to a larger
velocity dispersion than an inertial frame because the former adds a
position-dependent velocity in quadrature to each observed radial
velocity.  A non-zero velocity centroid suggests that the ensemble is
moving relative to the observer.  Figure 2 shows the distribution of
the velocities of the HVCs in the LSR and GSR frames of reference; the
latter are much smaller and suggest that the Galactic Center is a
better inertial frame than the Local Standard of Rest.  The mean
velocity of --46 $\pm$ 7 \kmse implies that the
Galactic Center is moving with respect to the barycenter of the
ensemble.  If we concentrate on the clouds with negative velocities
seen in the lower panel of Figure 3, the centroid of these clouds is
close to the Local Group barycenter ($l$ = 147\deg, \b = --25\deg,
$v_r$
= --82 \kms).  Relative to the LSR, the mean velocity of this group of
clouds is --173 \kms, relative to the GSR, it is --88 \kms.
However, relative to a frame of reference centered on the barycenter
of
the Local Group, the LGSR, the mean velocity is only --28 $\pm$ 10
\kms.
These numbers suggest that the barycenter of the Local Group is
the proper inertial frame for the HVCs and that the Milky Way is
approaching the barycenter at a velocity of about 60--90 \kms.

\begin{figure}[!htb]
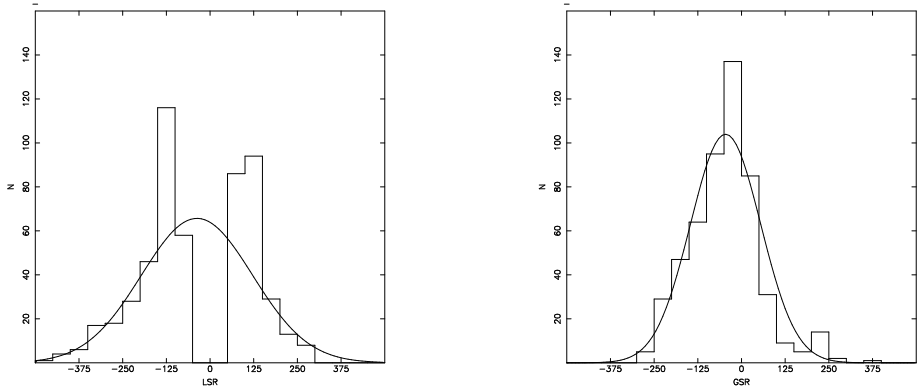

\hbox{
\psfig{figure=hvclsr.ps,height=2.0in,angle=0,silent=1}
\hspace{0.7in}
\psfig{figure=hvcgsr.ps,height=2.0in,angle=0,silent=1}}
\vskip -0.05in
\caption{ {\it Left:} Histogram of the distribution of HVC velocities
relative to the LSR.  HVCs which might have  a V$_{\rm LSR}$ near
zero would not be separable from
conventional--velocity Galactic emission. {\it Right:} Distribution of
HVC velocities relative to the GSR.  A Gaussian profile was fit to the
wings of both histograms: the GSR distribution of the HVCs is
more narrowly confined than that relative to the LSR, 
suggesting that the GSR system is the more appropriate inertial 
reference frame.  The data in both panels are from the Wakker \& van
Woerden (1991) catalogue of HVCs [from Blitz et al. (1999a)].}
\end{figure}

Recently, Burton \& Braun (this volume) and Braun \& Burton (1999)
have
suggested that there is a separate class of HVCs which they call
``Compact HVCs" or CHVCs that are different from those discussed and
analyzed above, but somehow share the kinematic properties of the
conventional HVCs.  In fact, the Burton \& Braun clouds are simply a
subset of the HVCs compiled by Wakker \& van Woerden (1991) with a few
additional clouds that fall between the latter's sampled points. Even
though the clouds listed in Burton \& Braun are chosen from the
smallish end of the Wakker \& van Woerden sample, the former have a
median surface area \lapprox 1.0 deg$^2$ compared to 1.5 deg$^2$ for the
Wakker \& van Woerden sample (Blitz et al. 1999), and are not 
more compact than the typical HVCs. Even this difference may result
largely because Braun \& Burton use the beam deconvolved area and
Blitz et al. (1999a) do not. In any event, the small difference
hardly warrants a new designation.  Apparently, Burton \&
Braun have succeeded in showing that a representative subset of HVCs
has properties of and behaves similarly to the HVC population as a
whole.

\subsection{Dynamical Simulation}

The Local Group is dynamically simple, thus it should be possible to
simulate the dynamical history of the HVCs and reproduce both the
spatial and kinematic distribution of the clouds.  Since 98\% of the
mass of the Local Group is in the Milky Way and M31, we modelled
 the
Local Group in Blitz et al. (1999a)
as a modified, restricted 3-body system with the HVCs as
test particles in a potential defined by the Galaxy and M31.  The
simulation begins with the HVCs on a regular grid and no initial
velocity dispersion.  The Milky Way and M31 are separated initially by
100 kpc and expand with the Hubble flow; M31 is taken to have twice
the
mass of the Milky Way.  Enough mass is put into the two galaxies so
that they are turned around from the Hubble flow until they reach
their
present separation and velocity of approach.  After the start of the
simulation, if a particle comes within 100 co-moving kpc of either
galaxy, we assume that the cloud is accreted.   This allows for a
somewhat larger interaction radius than the geometric cross-section of
the galaxies to allow for gravitational focusing, tidal disruption
and dynamical friction.

The basic results of the model are insensitive to the exact value of
the relative masses of the two galaxies, or the accretion radius.  The
model is exceedingly simple and contains no hydrodynamics (the test
particles are non-interacting), though it is not self-consistent in
that all of the mass is placed in the two galaxies at t=0 and
accretion
does not increase the masses of the galaxies.  These shortcomings 
are compensated by the simplicity of the model.
It contains no free parameters and no fine tuning is done to improve
the
comparison with the observations.

\begin{figure}[!htb]
\vskip -0.8in
\centerline{\psfig{figure=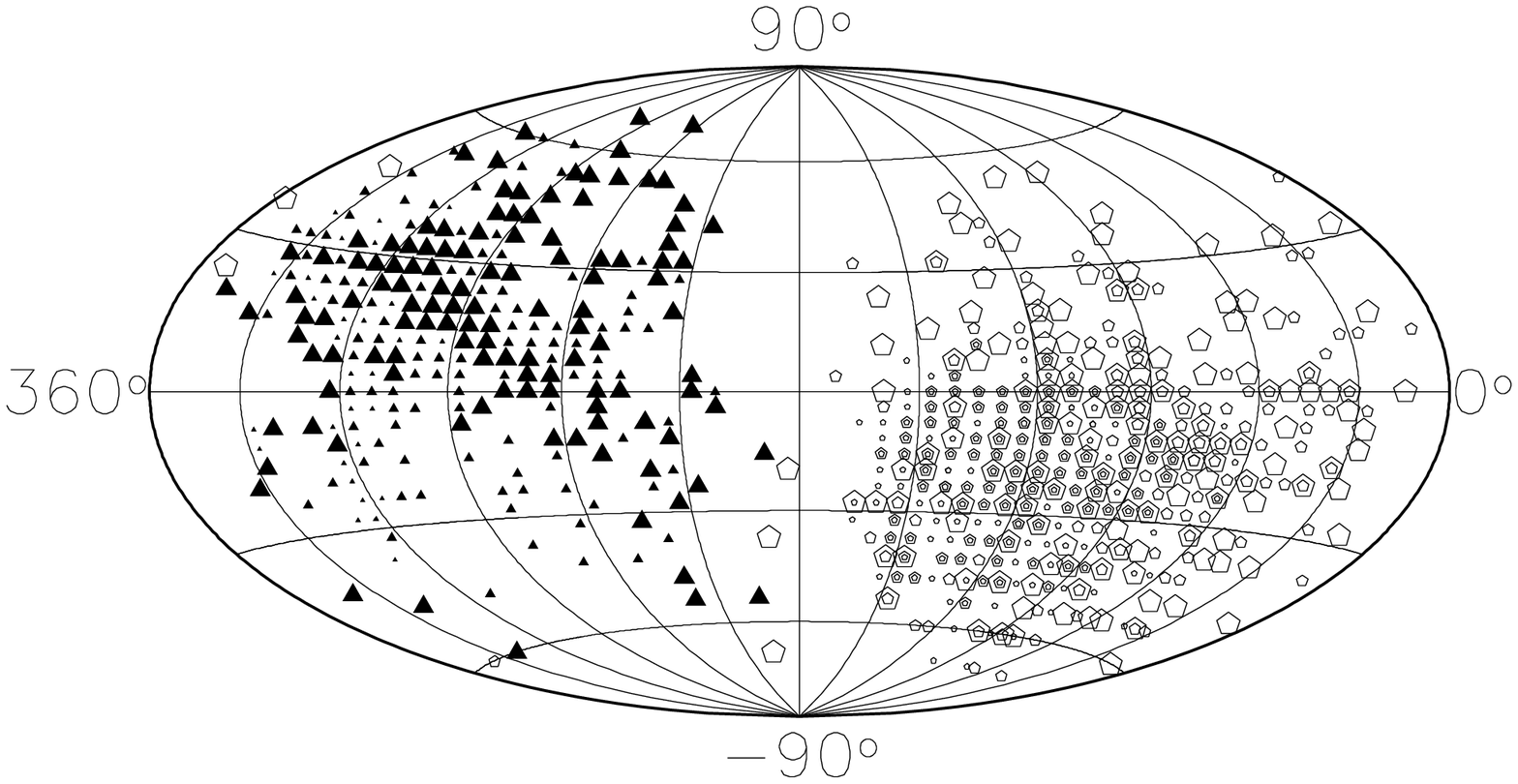,width=4in,angle=0,silent=1}}
\vskip -1.6in
\hbox{\hskip 0.24in
\centerline{\psfig{figure=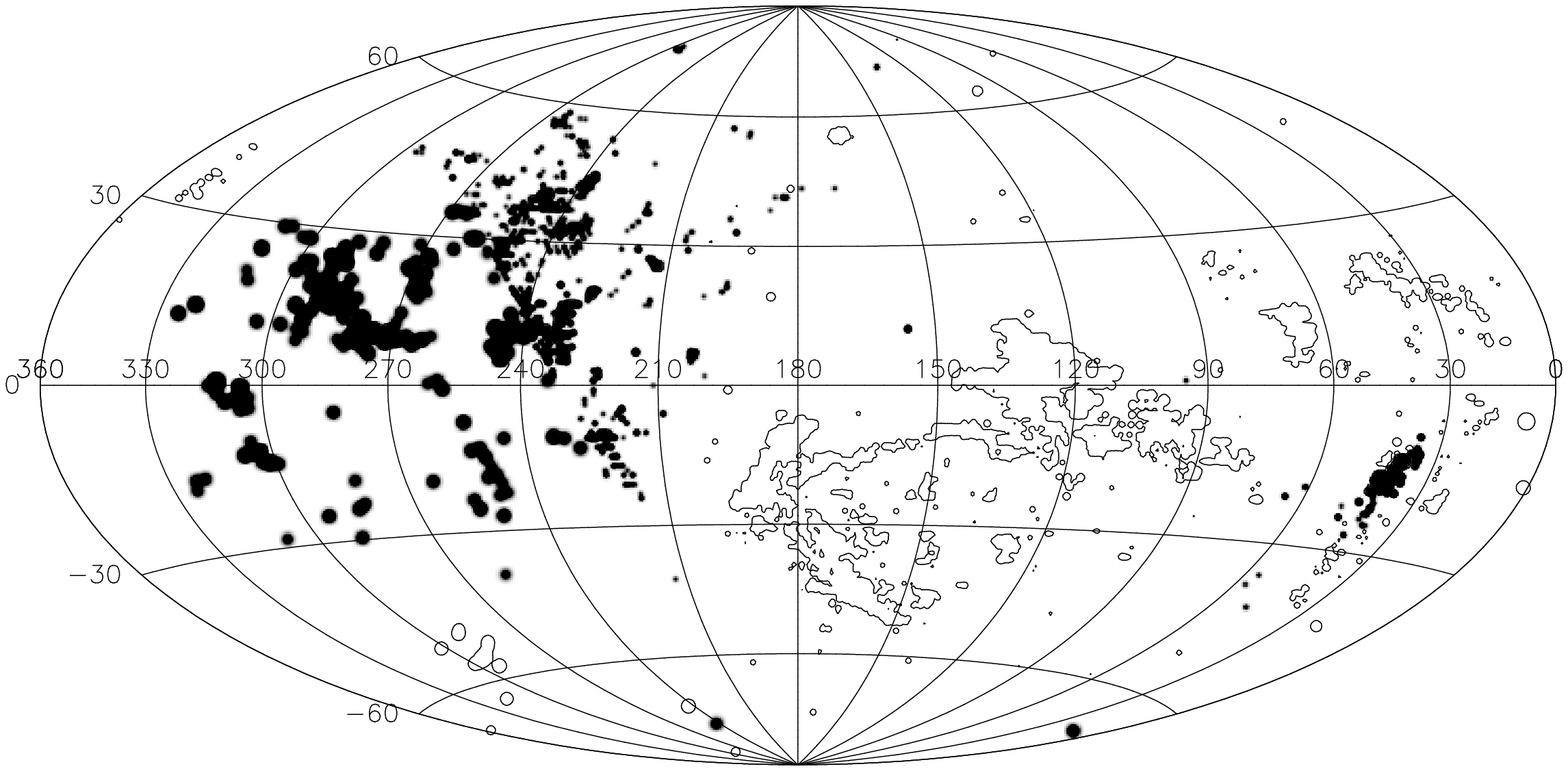,width=4in,angle=-90,silent=1}}}
\vskip -0.2in

\caption{Comparison of simulated and observed 
distributions of the HVC ensemble on the sky. {\it Upper:}
Distribution of all
simulated clouds having HI column densities greater than 3$\times
10^{18}$ cm$^{-2}$ and $|{\rm V}_{\rm LSR}|$ greater than 100 \kms.  The
size of the symbols is proportional to column density and ranges 
between $3\times10^{18}$ to greater than $3
\times 10^{19}$ cm$^{-2}$.  Strictly speaking, these simulated column
densities are total ones, i.e. including the dark--matter content.
The
triangles represent clouds with positive LSR velocities;  the
pentagons, clouds with negative LSR velocities.  This figure
represents
the distribution of HVCs if the clouds have not been destroyed by
passage through a hot intergalactic medium and if collisions between
HVCs are rare.  {\it Lower:}  Distribution of observed HVCs but
excluding the Magellanic Stream and the Northern Hemisphere complexes
A, C, and M, which are evidently relatively nearby and thus not
representative of the angular size of individual clouds in the
Local--Group ensemble. Positive LSR velocities are denoted by filled
contours, negative LSR velocities by open contours.  The lower panel
was kindly provided by Bart Wakker.[from Blitz et al. (1999a)]}
\end{figure}

The results are shown in Figures 3 and 4.  Figure 3 is a comparison of
the simulated and observed spatial distributions of the HVCs 
with the Magellanic Stream and the A, C and M complexes
removed.  The comparison shows a rather good agreement considering the
simplicity of the model.  The model reproduces the two concentrations
of clouds, the separation into positive and negative LSR velocities
and
the tilt in the positive and negative velocity cloud groups relative
to
the Galactic Plane.  No extraneous groups of clouds are produced.  No
other model considered to date reproduces all of these features of the
HVC spatial distribution.  The separation into two groups occurs
naturally 
in the model because the clouds are distributed along a wide filament
along
the line connecting the Milky Way and M31.  The negative LSR velocity
clouds are seen along the filament towards the Local Group barycenter,
and the positive LSR velocity clouds are seen primarily in the
antibarycenter direction.  Both groups are falling toward the LGSR.

\begin{figure}[htpb]
\hbox{
\psfig{figure=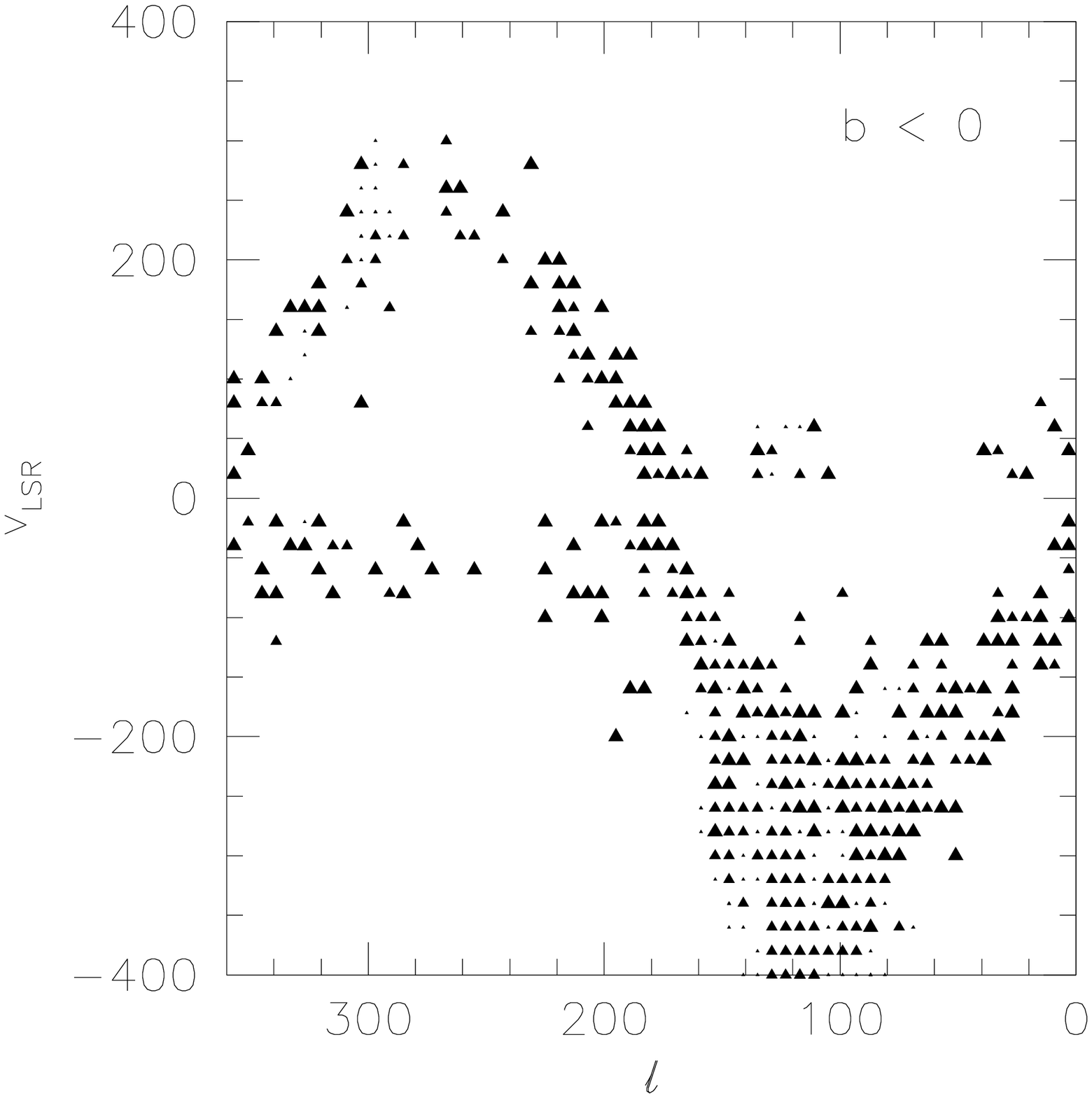,height=2.5in,angle=0,silent=1}
\psfig{figure=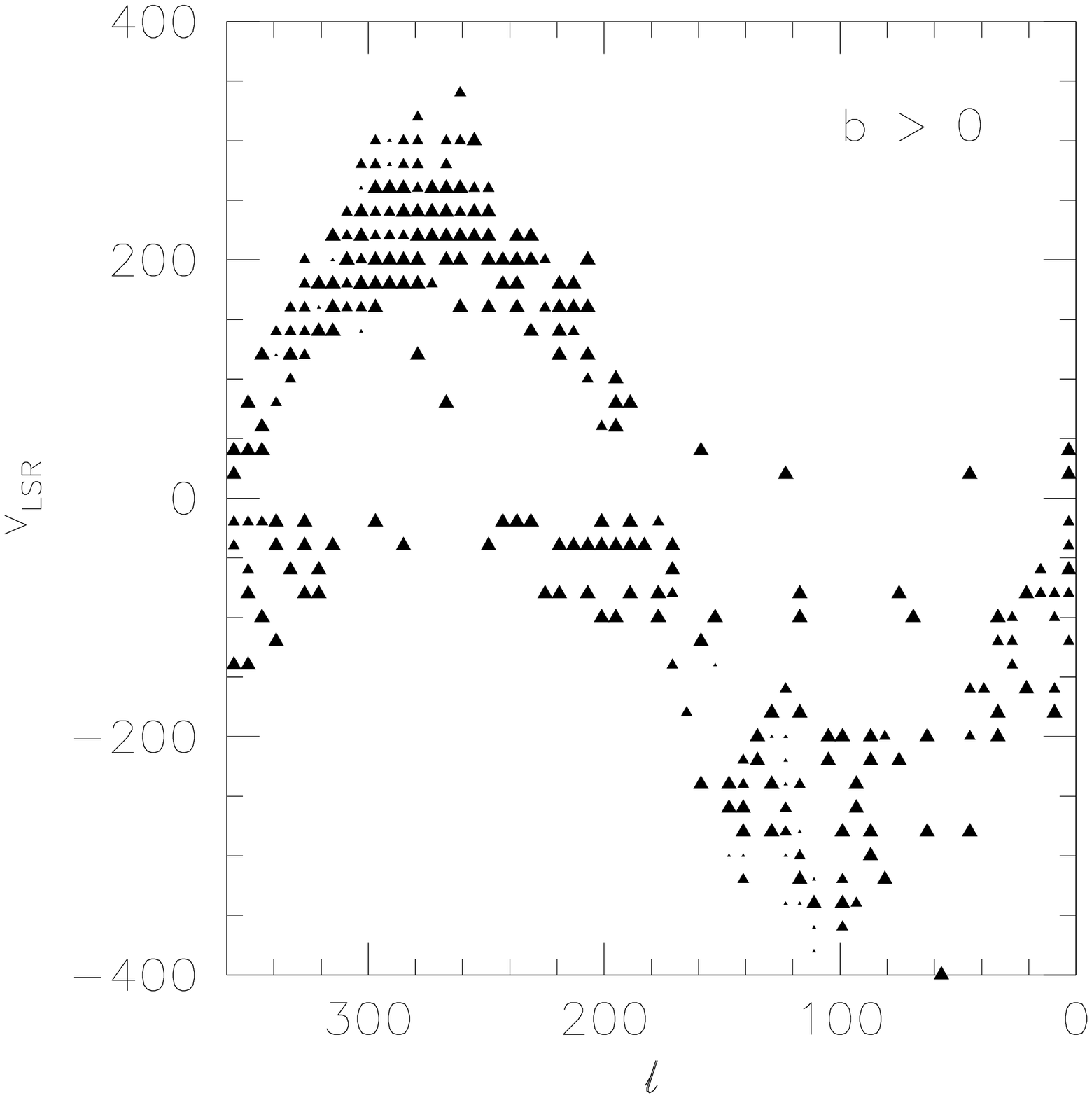,height=2.5in,angle=0,silent=1}}
\vbox{\vskip -0.1in}
\centerline{\psfig{figure=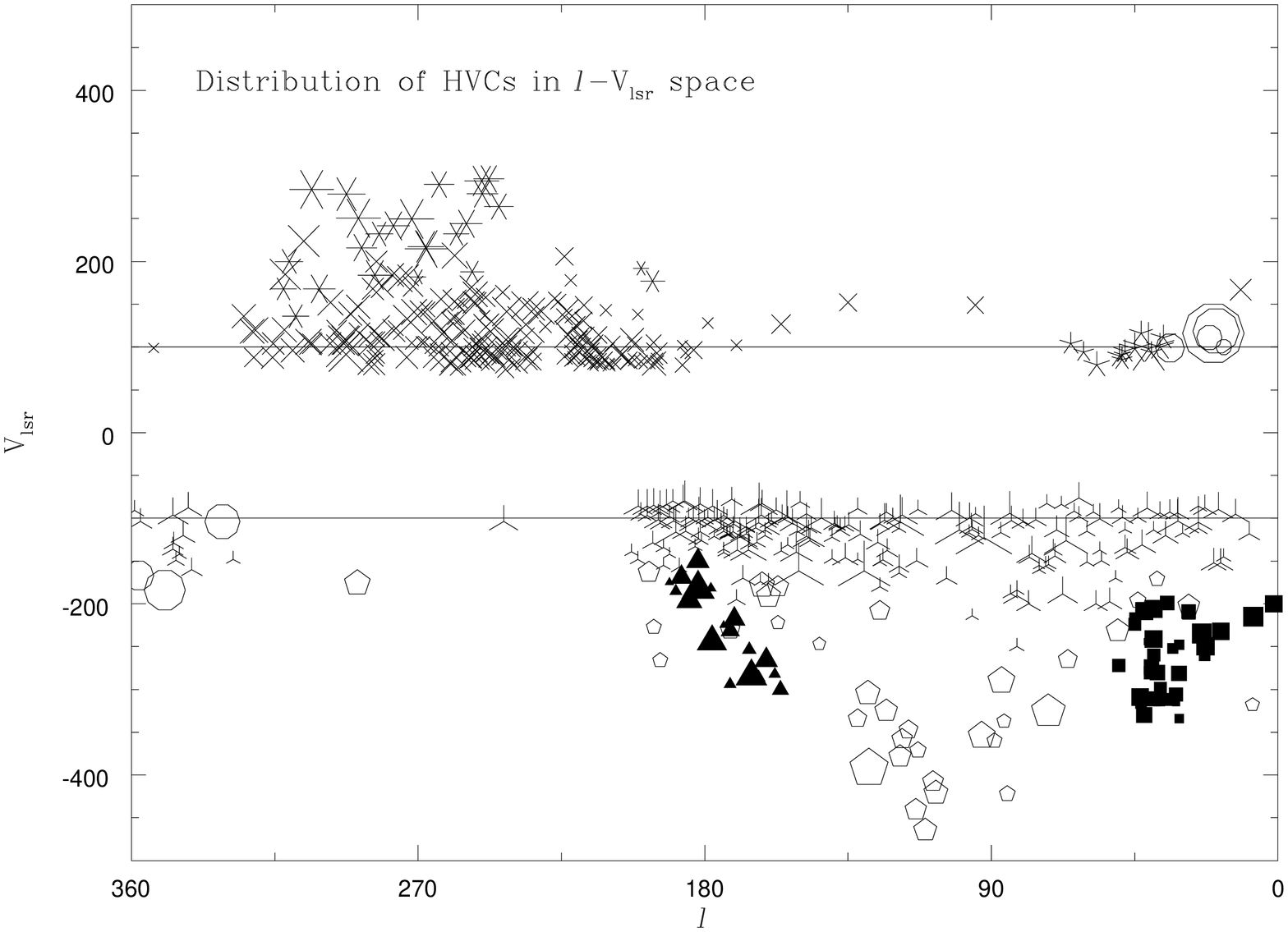,height=3in,angle=-90,silent=1}}
\vskip -0.2in
\caption{Comparison of the simulated longitude--velocity distribution
of the HVCs with the observed situation.  Radial velocities are
relative to the LSR.  {\it Upper:} Simulated kinematic distribution of
clouds with HI column densities greater than $3\times 10^{18}$
cm$^{-2}$, plotted separately for $b < 0 \deg$ and for  $b > 0 \deg$.
{\it
Lower:} Longitude--velocity diagram of the observed HVC ensemble, as
compiled by WvW91.  The symbols are proportional in size to the flux
from the individual clouds, and are keyed to the individual complexes
defined by Wakker (1991).  The MC, OA and A,C, and M Complexes are not
included in this Figure (see text).  Clouds with LSR velocities
$|{\rm V}_{\rm LSR}| < 80$ \kmse are not considered here as HVCs, regardless
of their location.[from Blitz et al. (1999a)]}
\end{figure}

The Magellanic Stream was removed from the comparison because it is a
group of clouds known to be of tidal orgin (Mathewson et al. 1974) and
is thus not well represented by our model.  Cloud C is by far the
largest cloud in the HVC ensemble and covers more than 1600 deg$^2$.
Clouds A and M, which are also quite large, have similar velocities
and may
be related to this large complex.  If the A, C and M complex is
gravitationally bound, it must be tidally unstable and thus quite
nearby.  The complex is also very elongated, consistent with tidal
shearing.  Thus, because of its apparent proximity and tidal shearing,
it is also not well represented by our dynamical model and is also
excluded from the comparison.

Figure 4 is a comparison of the longitude-velocity plot of the
observations and the simulations.  Again, the Magellanic Stream and
the
A, C and M Complex are removed.  The simulations reproduce the
sinusoidal envelope of the cloud ensemble, the offset toward negative
LSR velocities (due to the motion of the Milky Way toward the Local
Group barycenter) and the magnitude of the envelope of the
distribution.  This agreement in the quantitative aspects of the
$l - v$ distribution are particularly noteworthy because of the
absence of free parameters in the model.

\section{Implications and Speculations}

If we accept the model at face value, it implies that the HVCs are
formed
with the earliest structures in the Universe and are the building
blocks from which the Milky Way and M31 formed.  The HVCs that we see
today would then be the leftover building blocks that have not yet
been 
accreted by either galaxy.  If the Local Group is not unique, it
suggests that structures similar to the HVCs are responsible for {\it
all} initial galaxy formation, though there would be large differences
in
how galaxy evolution proceeds depending on the density of the
environment (see below).

If the HVCs are almost as old as the Universe, they must be
gravitationally bound and tidally stable.  If they have typical
distances of 1 Mpc as the model suggests, then the observed angular
sizes and velocity dispersions imply that to be self-gravitating,
about
90\% of the matter in the HVCs must be dark; the dark matter may be
either baryonic or non-baryonic.  Table 1 gives mean
derived parameters for the HVCs.  For example, the clouds could have a
90\%
ionization fraction, in which case the emission measure would be
$\simeq
10^{-2}$ cm$^{-6}$ pc, an undetectable value at present.  if
the dark
matter is the same as that in the halo of the Milky Way and M31, the
ratio of
luminous to dark matter is comparable in the HVCs and the galaxies,
just what one would expect if the Milky Way and M31 were assembled
from
HVCs.  In this case, the HVCs would also have characteristic masses of
$\simeq 10^8$ \msun, similar to the mini-halos
postulated by Ikeuchi (1986) and Rees (1986) to be the first
structures to form
after recombination.

\begin{table}
\caption{Mean Derived HVC Properties [from Blitz et al. (1999a)]}
\begin{center}
\begin{tabular}{lr}
\hline\hline
Quantity & Value \\
\hline
HI mass &  1.9 $\times 10^7$ \msun \\
Total Neutral Gas Mass & 2.7 $\times 10^7$ \msun \\
Total Mass    &  2.8 $\times 10^8$ \msun \\
Diameter & 28 kpc \\
Distance & 1 Mpc  \\
$n_{\rm HI}$ &  $0.7 \times 10^{-4}$ \cc   \\
\hline\hline\\
\end{tabular}
\end{center}
\vskip -0.35in
\end{table}

Our dynamical model allows us to calculate a mass accretion history, which
is shown in Figure 5.  The present day mass accretion rate is
estimated
to be 0.8--1.2 \msune yr$^{-1}$, approximately what is needed to fuel
the present day star formation in the Milky Way (e.g., Blitz 1995).
The orbits of some of the HVCs are likely to cross in the region
between M31 and the Galaxy, giving rise to collisions between the
HVCs.  Typical collision velocities can be estimated from Figure 2 to
be $\simeq$ 200 \kmse possibly giving rise to an x-ray halo
surrounding the
Milky Way and M31.  This gas would probably have a temperature of
about
10$^6$ K, and might be detectable.  To predict whether or not such an
x-ray halo
exists, and how large it would be requires adding 
hydrodynamics to our simulation.  The cooling time for the gas would
be
$>$ 10$^{10}$ yr.  Collisions between clouds could also be the source
of the
x-ray halos around poor groups and in denser extragalactic
environments.

\begin{figure}[!htb]
\centerline{\psfig{figure=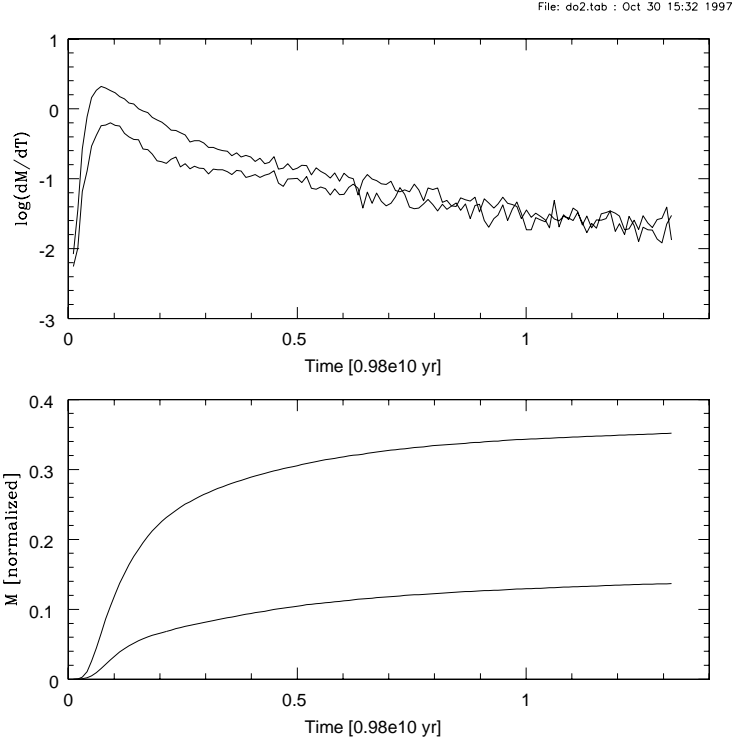,height=4in,angle=0,silent=1}}
\vskip -0.2in
\caption{{\it Upper:} Normalized rate of simulated accretion of
clouds for M31 (upper line) and for the Milky Way (lower line).
The M31 accretion rate is typically about twice that of the Milky Way.
After about $3\times 10^9$ y, the accretion rate becomes nearly
exponential with an e--folding time of about $5 \times 10^9$ y.  At
the
current epoch, the accretion rate is flattening out, and is equivalent
to about 7.5 \msune y$^{-1}$ (about 10 times that of the HI alone)
for the Milky Way.~~~{\it Lower:} Normalized
accreted mass for M31 (upper line) and for the Milky Way (lower line)
The plot shows that most of the mass is accreted at early times
and that additional mass is being added to both galaxies quite slowly
at the current epoch.  }
\end{figure}

In our picture, the growth of the Galactic disk is fueled
by the gradual accretion of HVCs and is consistent
with numerical simulations. Figure 6, for example, shows the results
of a hydrodynamical simulation in which a cloud similar to Complex
H is being accreted by a disk galaxy similar to the Milky Way.  The
gas streamer shown in the simulation is similar to that seen in
higher contrast versions of Figure 1.
Our picture of Milky Way formation is thus more consistent with the episodic 
accretion model of Searle and Zinn (1968) 
than it is with the Eggen, Lynden-Bell and
Sandage (1962) model.  Episodic evolution, furthermore, would lead to
metallicity correlations consistent with trends
seen in disk stars (Edvardsson et al. 1994).

\begin{figure}[htpb]
\centerline{\psfig{figure=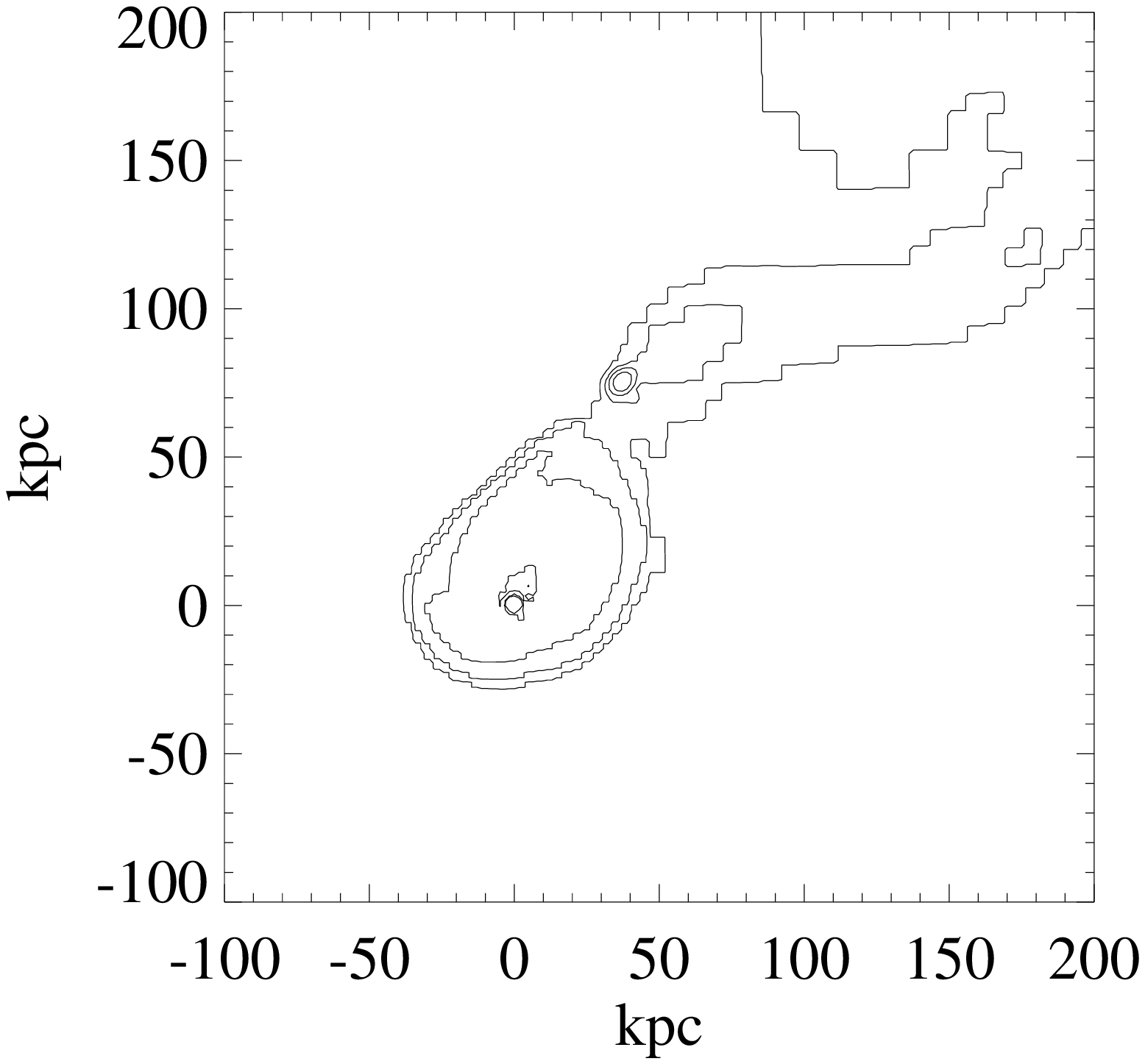,height=3.0in,angle=0,silent=1}}
\caption{This figure from Kepner (1998) shows the projected hydrogen
distribution
in a numerical simulation in a disk galaxy in a numerical simulation
by
J. Kepner and G. Bryan.  In the AMR hydrodynamics simulation of
a $\Lambda$-dominated CDM universe, they focused their high
resolution mesh on a binary galaxy system with properties similar
to the Milky Way.  In their simulation, the galactic disks
are built up primarily by the gradual accretion of gas clouds
with properties similar to those deduced for the HVCs.  In this
time-slice,
the main disk is accreting a gas cloud with properties similar to
Complex H.
The accreted cloud is being ram-pressure stripped by the hot gaseous
halo and is falling towards the disk.  Since it has relatively
high angular momentum, it eventually settles at the edge
of the preexisting disk.}
\end{figure}

Finally, the HVCs might be the z=0 analogues of the Ly$\alpha$ 
absorbing systems.  If the HVCs are indeed ubiquitous in the Universe,
they would correspond in column density to the Lyman limit systems.  
The frequency distribution of the Ly$\alpha$ absorbers is a power law
with a slope of --1.4 over 8 orders of magnitude in column density
(Wolfe 1993).  The
slope of the frequency distribution of column densities in the HVCs
has the 
same value of --1.4.  

\section {Predictions and Comparison with other Observations}

In our original paper (Blitz et al. 1999a), we made several
predictions based on our model.

\begin{enumerate}
\item  The HVCs should have substantially sub-solar metallicities.
These are not expected to be zero, since no extragalactic gas has 
primordial abundances.

\item  The HVCs should have low internal pressures, inconsistent with
a
Galactic origin. 	If the clouds are self-gravitating, then the
internal pressure
within the clouds can be given by 

\begin{equation}
P/k = 3\pi \alpha G \Sigma^2/20 k
\end{equation}

\noindent
where $\Sigma$ is the gas surface density, $\alpha$ = 2 for
self-gravitating clouds, and k is Boltzmann's constant (Bertoldi \&
McKee 1992).  For self-gravitating HVCs bound by their HI alone and a
surface density of $3 - 30 \times 10^{18}$ \c2, the expected mean
hydrostatic pressure within a cloud is expected to be 0.016 -- 1.6 K
\cc.  If the cloud is in a dark matter potential with 10 times the HI
mass, as we expect, the internal pressure would be about 10 times
higher, thus pressures of the order of 0.1 -- 10 K \cce are expected.

\item  There should be extragalactic analogues of the HVCs in other
extragalactic systems. 

\item  There should be H$\alpha$ emission associated with the HVCs
at a level at least as great as that detected toward the clouds in the
Magellanic Stream if the HVCs are Galactic in origin.  If they are
extragalactic, the emission measures should be \lapprox 0.1 cm$^{-6}$
pc.
\end{enumerate}

\noindent
In the past few months, several groups have reported new observations
that are consistent with our predictions and appear to contradict
the models with a Galactic HVC origin such as the Galactic fountain model:
\begin{enumerate}
\item Wakker et al.
(1999)  recently reported  a measurement of
sub-solar metallicity on a line of sight toward Mrk 290 in Complex C.  
They detected  SII, a species in the dominant ionization state and
which is not
depleted onto grains.  Wakker et al. obtained values for both the
atomic
and ionized gas components and concluded that the abundance along this
line of sight is only 0.094 solar.  They concluded that this cloud
represents an accretion event of an extragalactic cloud, in agreement
with 
our predictions.  Complex C has by far the largest angular
size of any HVC, and is probably the nearest.  If this cloud is of
extragalactic origin, it suggests that the other smaller clouds are
also
extragalactic.

\item Sembach et al. (1999) observed the ionized edges of HVCs in
the direction of Mrk 509 and PKS 2155-304 and detected strong C IV
absorption, with little or no CII or SiII.  The authors concluded that
the clouds are low density (n$_{\rm H}$ $\simeq$ 10$^{-4}$ \cc), large
(greater than several kiloparsecs) clouds with P/k $\simeq$ 2 K \cc.
This pressure is just in the range expected for self-gravitating,
dark matter confined HVCs, is four orders of magnitude less than
the pressure in the midplane of the Milky Way and two orders of
magnitude less than expected in the Galactic halo (Wolfire et al
1995). The density is also 
in good agreement with the derived value given in Table 1.

\begin{table}
\caption{Extragalactic HVC Analogues}
\begin{center}
\begin{tabular}{lrrr}
\hline\hline
Mass & Diameter & Galaxy & Reference \\
\hline
\msun               & kpc &  &  \\
\hline
$1.6 \times 10^8$ & 16 & M101 & van der Hulst \& Sancisi (1988)\\
$1.2 \times 10^7$& 5 & M101 & van der Hulst \& Sancisi (1988)\\
$1 \times 10^8$ &25& NGC 5668 & Schulman et al. (1996) \\
$5 \times 10^7$ & 7 & UM422C & Taylor et al. (1995;1996a)\\
$1.6 \times 10^8$ (?) & 16 (?) & UM 456B & Taylor et al. (1995;1996a)
\\
$1.4 \times 10^8$ (?) & 8 (?) & F495-IVB & Taylor et al. (1996b) \\
$7.9 \times 10^7$& 38 & NGC 628 & Kamphuis \& Briggs (1992) \\
$9.5 \times 10^7$& 47 & NGC 628 & Kamphuis \& Briggs (1992) \\
$2.1 \times 10^8$& 6  & NGC 3227 & Mundell et al. (1995)\\
$2\times 10^7$ & 28 & Local Group & This paper; Table 1 \\
\hline\hline\\
\end{tabular}
\end{center}
\vskip -0.5in
\end{table}

\item Extragalactic analogues of the HVCs have been seen toward a
number of
galaxies.  A list of such clouds is given in Table 2.  These clouds
were found serendipitously in the course of mapping other objects.
Many are seen in projection against other galaxies, some are seen as
distinct objects separated in both position and velocity from the
parent galaxy. Such clouds would appear as HVCs if viewed from the
target galaxy.  Clearly numerous extragalactic HI clouds have been
found with properties similar to those of the HVCs given in Table 1.
\smallskip

Several blind surveys of HI have been undertaken, notably by Zwaan
et al. (1996), and more recently by Spitzak \& Schneider (1999).  In
the Zwaan et al. survey, no extragalactic analogues were found without
optical counterparts.  Spitzak \& Schneider found one cloud without
an optical counterpart. Neither survey is particularly sensitive to HI
masses
typical of what we expect from extragalactic 
HVCs, though a few probably should have been detected in the
Zwaan et al. survey if it is as sensitive as was claimed.  It is
difficult to predict how many HVC analogues should be detected since
the number associated with a galaxy group or cluster probably depends
sensitively on the density of the environment. It is therefore
difficult to assess whether the non-detections in the blind searches
(except for Spitzak \& Schneider) are significant. A targeted,
high sensitivity survey in the direction of a good Local Group
analogue might well decide this issue.

\item
H$\alpha$ has been detected toward Clouds A, C and M (Tufte et al.
1998)
as well as toward the Magellanic Stream (Weiner \& Williams 1996).
There is some question of whether the emission toward
the Magellanic Stream is due to photoionization from the Galactic
ionizing radiation that leaks out of the plane of the Milky Way, or is
due to shock heating from the clouds as they pass though the diffuse
gas in the Galactic Halo. However, regardless of what produces the
H$\alpha$, HVCs that are of Local Group origin should
have lower emission measures than those detected toward either cloud
complex.  The H$\alpha$ measurements are a critical test of the Local
Group model and observations toward the very high velocity clouds
($|v_r| > 200$ \kms) should give emission measures no higher than
$\sim$ 0.1
cm$^{-6}$ pc.  Measurements are currently underway by the Wisconsin
group using their WHAM instrument, by a Maryland--Carnegie group 
and by a group in Australia.  If the HVCs are Galactic
(distances $<$ 50 kpc), their H$\alpha$ surface brightnesses should be
at least as large as those already detected.  Results should be
available within the next year.

\end{enumerate}

The low metallicity, pressure and density detected along
several lines of sight support a Local Group origin.  The
detection of extragalactic HI clouds with properties similar to those
inferred if the HVCs are Local Group objects suggests that such clouds
do exist in intergalactic regions. The relative paucity of HVC
analogues seen in blind HI surveys may simply be a result of
insufficient sensitivity or sky coverage, and more sensitive
observations perhaps directed toward poor galaxy groups might usefully
be undertaken.  H$\alpha$ measurements will provide a critical test
which
should be able to distinguish between Galactic or extragalactic 
locations for the HVCs.

\section {Other Possible Origins}

	One possibility other than a Local Group origin  
for the HVCs is that the clouds are extensive tidal debris from either
previous passages of the Magellanic clouds or other nearby
dwarfs.  Bland-Hawthorne et al. (1998) have recently suggested
that the so-called ``Smith Clouds'' are related to tidal streaming
associated with the passage of the Sgr dwarf (Ibata, Gilmore \&
Irwin 1994).
However, an important constraint on tidal models is the crossing time
for HVCs which can be written as follows:

\begin{equation}
t_{\rm c} ~~ = ~~ 17.1 ~ {{\Omega^{1/2} r_{\rm kpc}}\over {\Delta
v}}~~ {\rm My,}
\end{equation}

\noindent
where r$_{\rm kpc}$ is the distance from the Galactic Center in kpc
and $\Delta$v is the FWHM of the HI line averaged over the cloud.
The mean value of $\Delta$v for the HVCs is 30 \kms, and the median
value of $\Omega$ is 1.5 deg$^2$ (Blitz et al. 1999).  Thus the 
crossing time for a typical HVC is about 1 Myr/kpc, or about 50 Myr at
the distance of the Magellanic clouds.  HVCs at that distance cannot
be gravitationally bound (except perhaps for Complex C, H and the
Anticenter Complex). Clouds at 50 kpc therefore double in size in a
crossing time which corresponds to a decrease
in density of an order of magnitude.  Thus HVCs from a tidal origin
should not be able to survive for more than 1 -- 2 crossing times.
The orbital time for the Magellanic clouds is about 200 Myr, far too
long for HVCs to have been the result of a prior passage.  The
Magellanic stream itself has a much larger $\Omega$ than the typical
HVC and thus has a longer crossing time.  Nevertheless, the 
Magellanic Stream is identified over only about 1/4 of the sky,
suggesting that even those HI clouds are destroyed in about 50 Myr.
No other dwarf companions are close enough to the Milky Way to have
produced the extensive tidal debris that would be necessary to explain
the HVCs.

The Galactic fountain model postulates that the HVCs are HI clouds 
which have condensed from gas expelled into the halo by supernovae and
stellar winds (Shapiro \& Field 1976; Bregman 1980).  It has long been
known that the Galactic fountain model cannot produce HI clouds with
radial velocities in excess of the circular speed of the Galaxy of
about 220 \kms.  Many HVCs have significantly larger velocities.
Furthermore, the recent evidence for low HVC pressures and densities,
the low metallicities, and the inability of the fountain models to
reproduce the observed features seen in Figures 3 and 4 make the
Galactic fountain untenable for the majority of HVCs.  Nevertheless,
even the Local Group model seems to produce insufficient numbers of
clouds at low LSR velocities (clouds near V$_{lsr}$ = 100 \kmse -- see
Figure 4).  Some of these clouds are not fully separated in velocity
from the main Galactic emission and may therefore yet be part of the
normal Galactic emission or a Galactic fountain phenomenon.  Just as
H$\alpha$ measurements become a good test of the Local Group origin
for
the HVCs, it may be that the lower velocity HVCs will be relatively
bright H$\alpha$ emitters. 




\end{document}